\documentclass[aps,twocolumn,amsfonts,showpacs,superscriptaddress]{revtex4-1}
\usepackage[dvipdfmx]{graphicx}
\usepackage{amssymb,amsmath,amsthm,booktabs,mathtools}
\usepackage{bm}
\usepackage{color}
\usepackage{tikz}
\usepackage{enumitem}
\usepackage{units}
\usepackage{braket}
\usepackage{amsmath}
\usepackage{cancel}

\newcommand\normalorder[1]{{:}\mkern1mu#1\mkern1.6mu{:}}

\begin{document}

\title{Unraveling two-photon entanglement via the squeezing spectrum of light traveling through nanofiber-coupled atoms}

\author{Jakob Hinney}
\altaffiliation[Present address: ]{Department of Electrical Engineering, Columbia University, New York, NY, USA}
\author{Adarsh S. Prasad}
\affiliation{Vienna  Center  for  Quantum  Science  and  Technology, TU  Wien-Atominstitut, Stadionallee  2,  1020  Vienna,  Austria}
\author{Sahand Mahmoodian}
\altaffiliation[Present address: ]{Centre for Engineered Quantum Systems, School of Physics, The University of Sydney, Sydney NSW 2006, Australia}
\email{sahand.mahmoodian@sydney.edu.au}
\author{Klemens Hammerer}
\affiliation{Institute  for  Theoretical  Physics,  Institute  for  Gravitational  Physics  (Albert  Einstein  Institute), Leibniz  University  Hannover,  Appelstra{\ss}e  2,  30167  Hannover,  Germany}
\author{Arno Rauschenbeutel}
\author{Philipp Schneeweiss}
\author{J\"{u}rgen Volz}
\affiliation{Vienna  Center  for  Quantum  Science  and  Technology, TU  Wien-Atominstitut, Stadionallee  2,  1020  Vienna,  Austria}
\affiliation{Department of Physics, Humboldt-Universit\"{a}t zu Berlin, 10099 Berlin, Germany}
\author{Max Schemmer}
\email{maximilian.schemmer@hu-berlin.de}
\affiliation{Department of Physics, Humboldt-Universit\"{a}t zu Berlin, 10099 Berlin, Germany}

\begin{abstract}
We observe that a weak guided light field transmitted through an ensemble of atoms coupled to an optical nanofiber exhibits quadrature squeezing. From the measured squeezing spectrum we gain direct access to the phase and amplitude of the energy-time entangled part of the two-photon wavefunction which arises from the strongly correlated transport of photons through the ensemble. For small atomic ensembles we observe a spectrum close to the lineshape of the atomic transition, while sidebands are observed for sufficiently large ensembles, in agreement with our theoretical predictions. Furthermore, we vary the detuning of the probe light with respect to the atomic resonance and infer the phase of the entangled two-photon wavefunction.  From the amplitude and the phase of the spectrum, we reconstruct the real- and imaginary part of the time-domain wavefunction. Our characterization of the entangled two-photon component constitutes a diagnostic tool for quantum optics devices.
\end{abstract}

\maketitle

Non-classical states of light are at the heart of quantum optics. Many experimental approaches for the generation of non-classical states of light  are based on strong coupling between photons and quantum emitters, e.g., making use of resonant enhancement of atom-light interaction via high finesse optical cavities \cite{birnbaumPhotonBlockadeOptical2005, dayanPhotonTurnstileDynamically2008, faraonCoherentGenerationNonclassical2008, reinhardStronglyCorrelatedPhotons2012, reisererQuantumGateFlying2014, volzNonlinearPhaseShift2014, hamsenTwoPhotonBlockadeAtomDriven2017,guerrero_quantum_2020} or employing collective response of strongly interacting Rydberg atoms \cite{peyronelQuantumNonlinearOptics2012, dudinStronglyInteractingRydberg2012, parigiObservationMeasurementInteractionInduced2012, maxwellStorageControlOptical2013,  baurSinglePhotonSwitchBased2014, thompsonSymmetryprotectedCollisionsStrongly2017, stiesdalObservationThreeBodyCorrelations2018, tiarksPhotonPhotonQuantum2019}. These approaches aim to maximize the interaction strength between atoms and photons to generate non-classical states of light.

In contrast, it has recently been predicted that light with non-classical signatures can be generated in a conceptually simple system consisting of $N$ two-level emitters \emph{weakly} coupled to a continuum of modes propagating in a one-dimensional waveguide and driven with a coherent laser field~\cite{mahmoodian_strongly_2018}. This occurs even when the atoms are driven far below saturation.
This has led to the observation of highly correlated states of light which can be tuned to exhibit antibunching or bunching by controlling the optical depth of the atomic ensemble~\cite{prasadCorrelatingPhotonsUsing2020}. At low input powers, the photon correlations arise from the two-photon component of the field, which can be written as a superposition of a separable part and an entangled part~\cite{hanschke_origin_2020-2}. Measuring the second-order correlation function of the field, as in~\cite{prasadCorrelatingPhotonsUsing2020}, probes the absolute magnitude of the entire two-photon component. However, it does not provide direct access to the relative phase and amplitude of the entangled and separable parts.

\begin{figure}[h!]
\centering
\includegraphics[width=\linewidth]{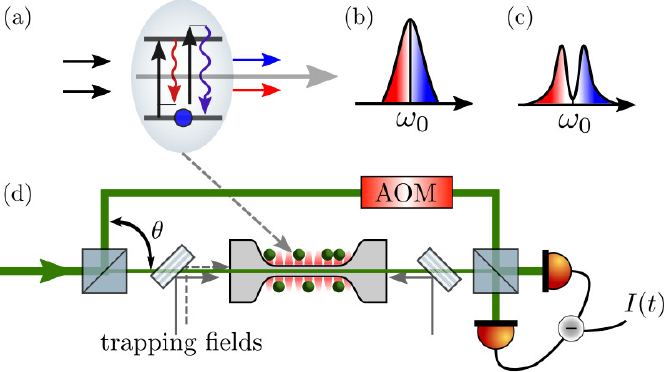}
\caption{(a) Atom-mediated photon interaction: Two photons interact with a two-level atom and exchange energy such that they become red- and blue-detuned with respect to the incident light. (b) The scattered light exhibits a Lorentzian shaped spectrum.
(c) As the generated photon pairs propagate through the ensemble, absorption around the atomic resonance $\omega_0$ attenuates the central part of the spectrum, leaving red- and blue-detuned sidebands. (d) Schematic setup: Probe light exiting the nanofiber interferes with a local oscillator at phase $\theta$ on a $50:50$ beam splitter and is analyzed on  balanced photo-detectors. Atoms are trapped in the evanescent field of the nanofiber-waist of a tapered optical fiber by a combination of red-detuned standing-wave light field at a wavelength of \unit[935]{nm} (solid gray line) and blue-detuned running-wave light field (dashed gray line) at \unit[685]{nm}.}
\label{fig_setup}
\end{figure}

Here, we report the observation of quadrature squeezing~\cite{slusher_observation_1985,lu_observation_1998-3, ourjoumtsevObservationSqueezedLight2011a, schulteQuadratureSqueezedPhotons2015, wallsReducedQuantumFluctuations1981} of the light that is transmitted through an ensemble of atoms coupled to a nanofiber. 
While the measurement of squeezing at low powers in a weakly coupled atomic ensemble is novel in itself, we also use this measurement to gain direct experimental access to the phase and magnitude of the entangled part of the transmitted field. More precisely, to leading order, the squeezing spectrum is proportional to the amplitude of the entangled part of the two-photon wavefunction, and the homodyne measurement of a continuous field allows us to measure the entanglement of the spectral components of opposite frequency, $\omega_0 \pm \omega$, around the probe field with frequency $\omega_0$~\cite{zippilliEntanglementSqueezingContinuouswave2015}.
The entanglement in the transmitted light is a consequence of the collective nonlinear response of the atomic ensemble. Entangled photons from each atom constructively interfere which leads to collectively enhanced squeezing.  
At the single atom level, the entanglement arises from the two-photon scattering process of resonance fluorescence, as depicted in Fig. 1(a). Two resonant photons that arrive during the lifetime of the excited state are scattered into spectrally entangled blue- and red-detuned sidebands, resulting in energy-time entangled photon pairs. This process resembles degenerate four-wave
mixing (FWM) by a Kerr non-linearity~\cite{carmanObservationDegenerateStimulated1966}. However, unlike FWM in a single-mode cavity QED \cite{Reiner2001JOSAB, reid_quantum_1986,Varada1987OPTCOMM}, the multimode nature of the cascaded system produces richer physics and requires a full multimode treatment of the light field. In our theoretical and experimental analysis we characterize the photon pairs by measuring the resulting quadrature squeezing. The latter allows us to infer the magnitude and the phase of the entangled part of the two-photon wavefunction.

We consider the physical setting of $N$ emitters weakly coupled to a single-mode continuum of the electromagnetic field with a continuous spectrum. In our setup, we use laser-cooled Cesium atoms in the evanescent field surrounding a tapered optical fiber with a \unit[400]{nm} diameter. The atoms are trapped in two 1D arrays of trapping minima along the nanofiber created through a combination of red- and blue-detuned fiber guided light fields~\cite{lekienAtomTrapWaveguide2004,vetschOpticalInterfaceCreated2010,corzo_large_2016-1}. The atoms are located at a distance of \unit[$\sim 250$]{nm} from the fiber surface, and each site contains at most one atom.
The nanofiber-guided probe field of power $P_\mathrm{in}$ is near-resonant with the Cesium D2-line transition and interfaces the atoms via the evanescent field of the nanofiber mode as depicted in Fig.~\ref{fig_setup}(d)~\cite{vetschNanofiberBasedOpticalTrapping2012}. The coupling of individual atoms to the nanofiber mode is weak, with a coupling constant $\beta = \Gamma_{\rm wg}/ \Gamma_{\rm tot} = 0.0070(5)$~\footnote{See Supplemental Material for the determination of $\beta$.}, where $\Gamma_{\rm wg}$ is the spontaneous emission rate into the waveguide and $\Gamma_{\rm tot} = 2\pi \times \unit[5.2]{MHz}$ is the total emission rate. We analyze the transmitted light via a balanced homodyne detection scheme: The output is first filtered from the trapping light fields and then sent to a 50:50 beam splitter where it is mixed with a local oscillator~(LO) field~\cite{das_measurement_2010,beguin_generation_2014-2,jalnapurkar_measuring_2019} as shown in Fig.~\ref{fig_setup} (d). The two outputs are measured on balanced photo-detectors, and we record the amplified differential current $I(t)$ between both photodiodes, which is proportional to the field quadrature $X_\theta(t)$. Here, $\theta$ is the relative phase between the LO and the probe field. The quadrature operator is given by $\hat{X}_{\theta}(t) = \frac{1}{2} \left[ \hat{a}(t)e^{i \theta} + \hat{a}^{\dagger}(t) e^{-i \theta}  \right] $, where $\hat{a}\,(\hat{a}^{\dagger})$ is the annihilation (creation) operator. For a classical (coherent) state of light $\langle \hat X_{\theta}(t)\rangle $ oscillates as $\cos(\theta)$, and the fluctuations $\Delta\hat{X}_{\theta}(t) = \hat{X}_{\theta}(t) - \langle \hat{X}_{\theta}(t) \rangle$ are independent of the phase $\theta$ and have minimum uncertainty, i.e., they are at the shot noise limit.

We probe an ensemble of $N$ atoms with an incident light that has a saturation parameter $s = P_\mathrm{in}/P_\mathrm{sat}$, where $P_{\mathrm{sat}} = \hbar\omega \Gamma_\mathrm{tot}/(8 \beta) = \unit[136 \pm 10]{pW} $, and detuning $\Delta$ from the atomic resonance. 
The transmitted light experiences a change in phase and amplitude upon interaction such that the output field has the average field quadrature
\begin{equation}
\langle \hat X_{\theta}(t)\rangle  =  |t_\Delta|^N  s^{\frac{1}{2}} \sqrt{\frac{ \Gamma_{\rm tot}}{8 \beta}} \cos(\theta + \delta) + \mathcal{O}\left(s \right).\label{eq_phase_shift_atoms}
\end{equation}
Here, $\mathcal{O}(s)$ refers to terms of order $s$ and higher, and $\delta = \mathrm{Arg}\lbrace t_\Delta^N\rbrace$, where $t_\Delta = 1 - 2\beta/(1- 2i \Delta/\Gamma_{\rm tot})$ is the single-photon transmission coefficient of a single atom. 
The atoms also modify the variance $\langle \Delta \hat X_\theta^2(t)\rangle$ and lead to quadrature squeezing. For squeezed light, the quadrature variance is smaller than the shot noise level for a certain value of $\theta$. More precisely, 
the variance $\langle \Delta \hat{X}^2_{\theta}(t)\rangle$  oscillates twice as fast as  $\langle \hat{X}_{\theta}(t)\rangle$ and exhibits a $ \cos (2 \theta + \varphi)$ modulation. For a resonant drive the phase offset $\varphi$ is zero and the light is amplitude squeezed with maximum squeezing observed for $\theta= (0, \pi)$.

\begin{figure}[!t]
\centering
\includegraphics[width=1.0\linewidth]{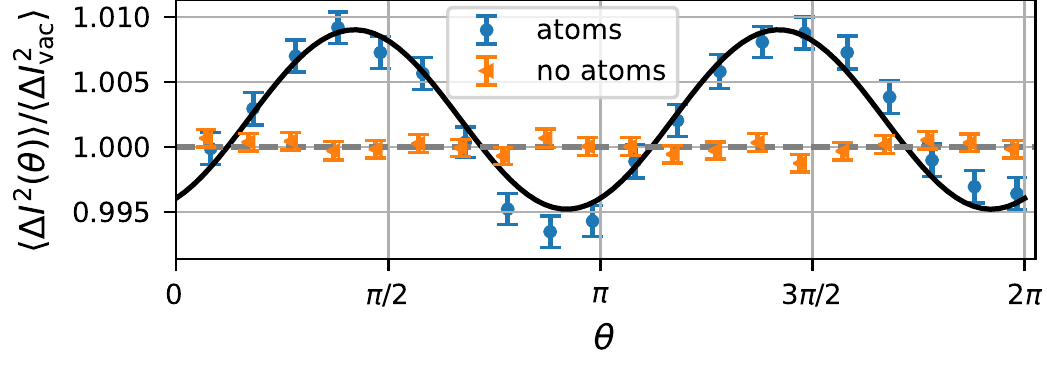}
\caption{The observed output noise as a function of the phase $\theta$. The noise is normalized to the vacuum reference and deduced from the frequency range $f_{\mathrm{min}}=\unit[1.5]{MHz}$ to $f_{\mathrm{max}}=\unit[23]{MHz}$. The signal obtained with trapped atoms (shown in circles) is compared to the measurement without atoms (shown in triangles). We fit the experimental data with trapped atoms using the function $ -A \cos(2\theta + \varphi) + c$ (black line) which reveals the expected $\cos(2\theta)$ modulation.
Squeezing occurs for $\theta $ around $0$ and $\pi$ while anti-squeezing occurs around  $\pi/2$ and $3\pi/2$. The resonant input power is $P_{\mathrm{in}}/P_{\mathrm{sat}}= 0.51 \pm 0.04$. }
\label{fig_theta_plot}
\end{figure}

In our setup, the squeezing is induced by the nonlinear optical response of atoms which behave as effective two-level systems. 
In the following, we limit our discussion to powers well below saturation $ s/8 \ll 1$ such that we can neglect events where three or more photons arrive simultaneously at any atom. 
The transmitted part of the two-photon wavefunction can be expressed in terms of separable and entangled photons. In the frame rotating with $\omega_0$, the two-photon wave function is,
\begin{equation}
    \psi_2(x_1, x_2) = t_{\Delta}^{2N} - \phi_N(x_1 - x_2).
\end{equation}
The first term denotes separable photons which are monochromatic traveling waves extended in space. Upon interaction with each emitter, each of the separable photons acquires a transmission coefficient $t_{\Delta}$. The entangled part of the wavefunction is not separable and is a localized function of the relative coordinate $x_1 - x_2$. On resonance ($\Delta =0$) and for a single emitter it is a decaying exponential 
\begin{equation}
\phi_{N=1}(x= x_1 - x_2) = 4\beta^2 e^{-|x|\Gamma_{\mathrm{tot}}/(2v_g) },\label{eq_phi_1}
\end{equation}
where $v_g$ is the group velocity of the photons. For $N$ emitters the entangled part of the wavefunction $\phi_N(x)$ was computed in Ref.~\cite{mahmoodian_strongly_2018}.
\begin{figure*}[!t]
\centering
\includegraphics[width=\textwidth]{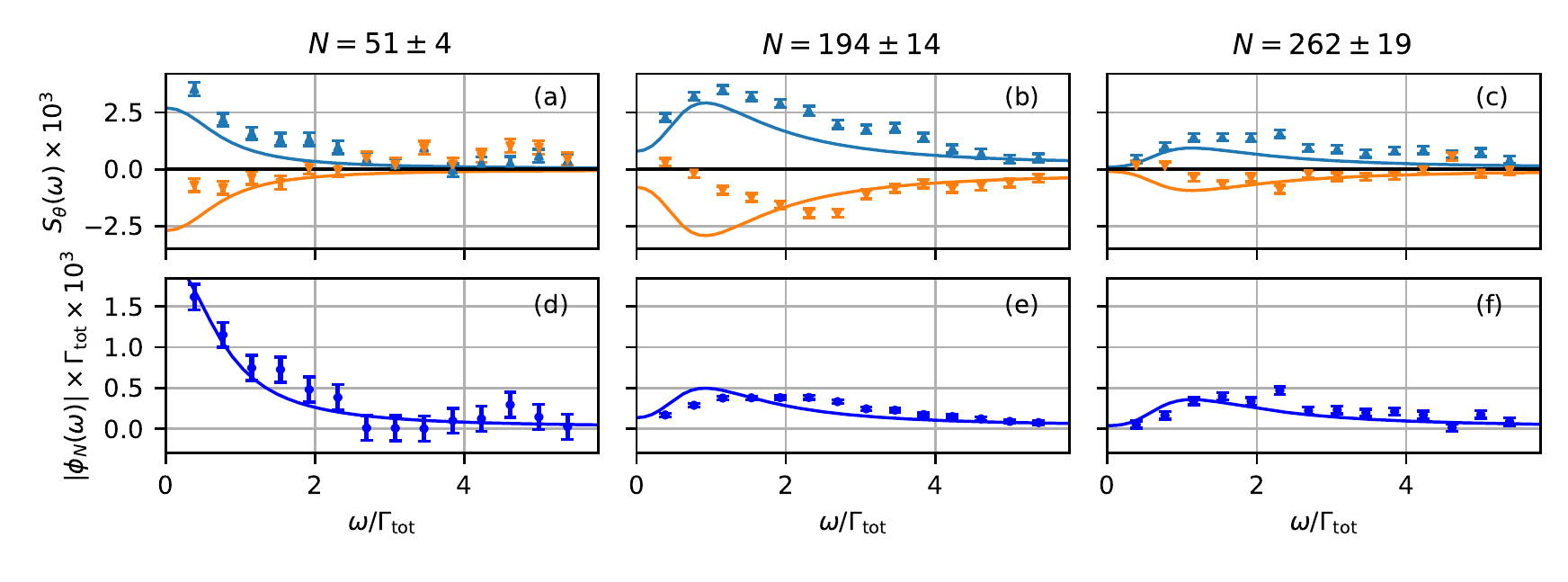}
\caption{The first row of panels, (a) -- (c), shows the squeezing spectrum $S_{\theta}(\omega)$ for three different atom numbers at input powers $s= (0.15\pm0.01 , 0.67\pm 0.05   ,0.29\pm0.02 $) (from left to right) at the angle of largest squeezing, $\theta=(0,\pi)$ in orange (bright) and largest anti-squeezing, $\theta = (\pi/2,3\pi/2)$ in blue (dark). The corresponding theoretical predictions are shown as solid lines. The second row of panels, (d) -- (f), shows the entangled photon spectrum $|\phi_N(\omega)|$ which is deduced from the upper row with the corresponding theoretical prediction (solid line).  
All theoretical curves are predictions based on the independently measured parameters $\beta$ and $N$, without any free fit-parameter taking into account the effect of independently estimated photon loss and detection efficiency in our setup~\cite{seesupplementalmaterialforadiscussion1,appel_electronic_2007,bachor_hansa_quantum_2019}.}
\label{fig_spectrum}
\end{figure*}

Introducing the normally ordered squeezing spectrum
\begin{equation}
S_{\theta}(\omega) = \int_{-\infty}^{\infty} \langle \normalorder{ \Delta \hat X_{\theta}(\tau)\Delta \hat X_{\theta}(0)} \rangle \, e^{i \omega \tau} \; \mathrm{d}\tau,
\end{equation} 
allows relating the entangled part of the two-photon wavefunction with the variance of the field quadrature~\cite{seesupplementalmaterialforadiscussion2}. Here, $\normalorder{\ldots}$ denotes normal ordering and the normally ordered squeezing spectrum of a coherent state yields $S_{\theta}(\omega) = 0$, while squeezed light yields $S_{\theta}(\omega) < 0$.
For the states generated in our experiment and in the case of weak saturation, the squeezing spectrum $S_{\theta}(\omega)$ and the spectrum of the entangled photons $\phi_N(\omega) =  \int \phi_N(x) e^{-i \omega x/v_g} dx$ are linked by
\begin{equation}
S_{\theta}(\omega) = -  \frac{  \Gamma_{\rm tot} }{16 \beta}  |\phi_N(\omega) | \cos\left[2\theta + \varphi_N (\omega) \right] s  + \mathcal{O}\left(s^2 \right),
\label{eq_spectrum}
\end{equation}
where we introduced the phase and the magnitude of the spectrum of the entangled two-photon wavefunction as $\phi_N(\omega) = |\phi_N(\omega)|e^{i\varphi_N(\omega)}$. In the following, we will first focus on a resonant probe field for which $\phi_N(\omega)$ is a real quantity ($\varphi_N(\omega) = 0$). For a single emitter $\phi_{N=1}(\omega)$ is the Fourier Transform of Eq.~\eqref{eq_phi_1} which gives a Lorentzian. Consequently, also the  squeezing spectrum has a Lorentzian shape~\cite{collettSpectrumSqueezingResonance1984}.
For $N$ emitters with $N\beta \ll 1$, re-absorption can be neglected and the scattered components constructively interfere since the process relies on forward scattering. 
A coherent build up of the squeezed photons takes place, and the squeezing spectrum is coherently enhanced, i.e. its amplitude  is $N$ times larger than the single atom squeezing spectrum~\cite{heidmannSqueezingManyAtom1985}. For large optical depth (OD), the probability that the squeezed photons are scattered again and thereby most likely removed from the fiber cannot be neglected anymore, and the problem becomes a quantum many-body problem. Recently, it has been shown that this problem can be solved exactly up to two-photon input states for chiral coupling where atoms couple only to one propagation direction of the mode~\cite{mahmoodian_strongly_2018}. Applying this formalism~\cite{seesupplementalmaterialforadiscussion2} allows calculating the squeezing spectrum for arbitrary $N$. 
The results from those calculations can be understood in the following manner: Photon losses occur predominantly  close to the emitters resonance which reduces the observed squeezing close to resonance. For many emitters this leads to a squeezing spectrum which develops sidebands due to the loss of squeezed photons that are resonant with the atoms as we will experimentally show later below.

We typically load a few hundred atoms into the evanescent field trap and determine the number of atoms $N$ in a separate transmission measurement. We then probe the atoms on the cycling transition of the D2-line. The probing lasts for $\unit[10 -  100 ]{\mu s}$, such that heating due to resonant scattering of the probe is small, and the number of trapped atoms does not change significantly during probing. After the homodyne measurement, we eject the atoms from the trap, shift the LO field frequency by $\unit[1]{MHz}$ and increase its power in order to observe a beat note between the probe field and the LO. From the beat note, we extract the relative phase $\theta$ between the probe field and the local oscillator at the moment of the homodyne measurement.
After this heterodyning stage, we switch off the probe field and record a vacuum reference where only the LO field is incident on the homodyne detector. 
We repeat the measurement $10\,000 - 100\,000$ times depending on the dataset and record $I(t)$ in each run. 
From each experiment cycle, we extract the power spectrum of $I(t)$ during atom probing, the vacuum reference, and the relative phase $\theta$. 

In a first step, we extract the noise $\langle \Delta I^2(\theta) \rangle$ and normalize it to the vacuum reference as shown in Fig.~\ref{fig_theta_plot}. We average the noise within the relevant frequency range of $f_\mathrm{min}= \unit[1.5]{MHz}$ and $f_\mathrm{max}=\unit[23]{MHz}$. The lower boundary is chosen to exclude technical noise and the upper boundary to capture the physically relevant frequency range on the order of a few $\Gamma_{\mathrm{tot}}$.
The mean atom number during probing consists of $N  = 169\pm12$ trapped atoms.
In all measurements, the number of atoms decreases by less than $20\%$ during probing, which we infer in a separate transmission measurement. The incident field is resonant and has a saturation parameter  of $s  = 0.51 \pm 0.04 $.
In Fig.~\ref{fig_theta_plot} data points with atoms show the expected $-\cos (2 \theta)$ modulation of the noise. Values smaller than 1 show that the light is quadrature squeezed.  The maximum observed squeezing within the bandwidth $\Delta f =  f_\mathrm{max} - f_\mathrm{min}$ is~$0.65\pm 0.12\,\%$. We fit the data with the function $-A\cos(2\theta + \varphi) + c$, where $A$ is the amplitude of the squeezing, and $c$ accounts for additional noise sources in the experiment. We obtain a small value for the squeezing angle $\varphi/\pi =  0.1 \pm 0.03$. This shows that the light is almost purely amplitude squeezed, i.e. the strongest squeezing is observed for $\theta$ close to $ 0$ and $\pi$.

In the next step, we make use of the homodyne detection scheme to  access the spectrum of the entangled photons. From the experimental power spectrum normalized to the vacuum spectrum, we deduce the  normally ordered squeezing spectrum $S_{\theta}(\omega) $~\cite{seesupplementalmaterialforadiscussion2}. Fig.~\ref{fig_spectrum} (a) -- (c) show the normally ordered squeezing spectrum for different $N$ for the most squeezed ($\theta =0,\pi$) in orange (bright) and the most anti-squeezed ($\theta =\pi/2,3\pi/2$) in blue (dark). Here, we average over a $\theta$-range of $\pm 18^\circ$ around both maxima and minima of the noise $\langle \Delta I^2(\theta) \rangle $. At $\theta = (0,\pi)$, for all atom numbers, the squeezing spectrum $S_{\theta}(\omega)$ exhibits fluctuations below zero which confirms that the spectral components created by the two-photon scattering are energy-time entangled~\cite{zippilliEntanglementSqueezingContinuouswave2015}.  
We attribute the deviations at low frequencies mainly to technical low-frequency noise.

In order to determine the spectrum of the entangled two-photon components $|\phi_N(\omega)|$ we make use of the fact that $S_{\theta}(\omega)$ is  proportional to $|\phi_N(\omega)|$ and for a resonant probe can be best extracted at the extrema of the cosine modulation ($\theta = 0,\pi/2,\pi,3\pi/2$).  We obtain $|\phi_N(\omega)|$ directly from the squeezing spectra shown in Fig.~\ref{fig_spectrum} (a) -- (c) by using Eq.~\eqref{eq_spectrum} and averaging over the absolute value of $S_\theta(\omega)$ at the four different values of $\theta$. 
Panels (d) -- (f) in Fig.~\ref{fig_spectrum} show the spectrum of the entangled photons $|\phi_N(\omega)|$ for different $N$ from small to large atom numbers.  For a small atom number, i.e., a small OD of the atomic ensemble, the spectrum is mainly dominated by a coherent built-up of entangled photons, and the shape of $\phi_N(\omega)$ is close to the lineshape of the atomic transition as shown in Fig.~\ref{fig_spectrum} (d). As $N$ is increased, as in (e) and (f), the probability that squeezed photons are scattered a second time and thereby lost from the waveguide mode increases. These events are more likely close to the atomic resonance. Consequently, one observes two sidebands, to the left and to the right of the atomic transition. Already for $N=194\pm14$ the spectrum strongly deviates from a Lorentzian. For $N= 262\pm19$, the entangled photons with frequency components close to resonance have almost vanished and  the entangled photon pairs are concentrated in the sidebands.

\begin{figure}
    \centering
    \includegraphics[width=\linewidth]{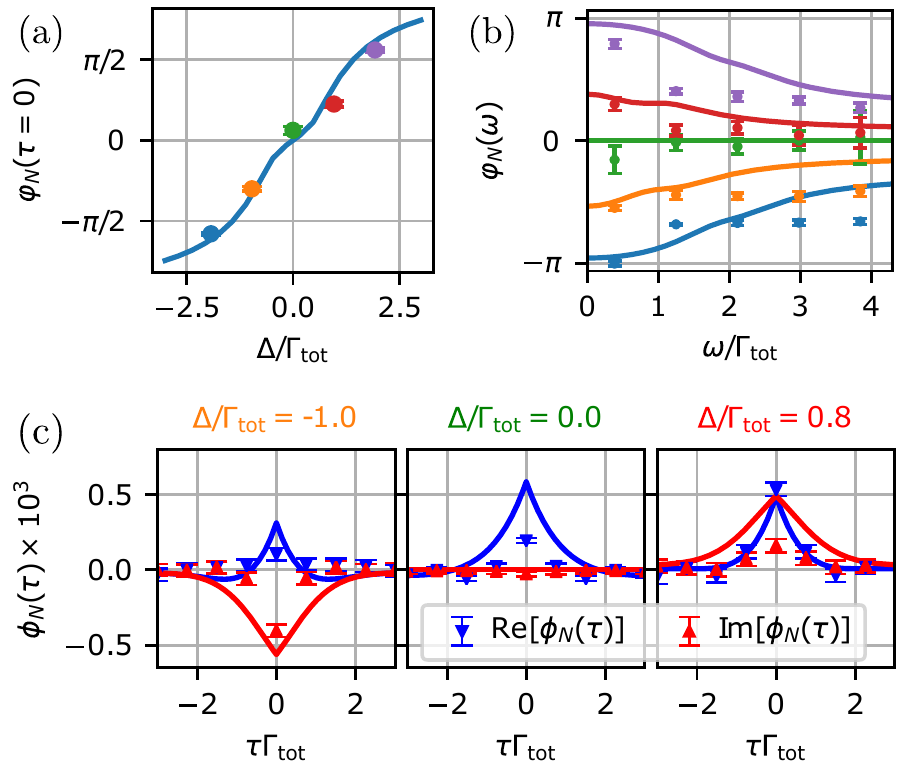}
    \caption{The phase of the entangled part of the two-photon wavefunction $\varphi_N$ for $N=140$ together with the corresponding theoretical predictions in solid lines. (a)  The integrated phase  $\varphi_N(\tau=0)=\mathrm{Arg}\lbrace \phi_N(\tau=0)\rbrace$ for different detunings $\Delta$.
    (b) The phase $\varphi(\omega)$ as a function of frequency. The detunings from top to bottom are $\Delta/\Gamma_\mathrm{tot} = 1.9,0.8,0,-1,-1.9$ and share the same color code as in (a). (c) The reconstructed real and imaginary part of the entangled two-photon time-domain wavefunction $\phi_N(\tau )$. The solid lines show the corresponding theoretical predictions based on the independently measured values of $\beta$ and $N$, without any free fit-parameter.}
    \label{fig_phase_detuning}
\end{figure}

So far, we considered a resonant probe field which leads to zero phase of the entangled part of the two-photon wavefunction ($\varphi_N(\omega) =0 $). In the following, we  measure the detuning dependence of $\varphi_N(\omega)$. We probe $N=140$ atoms with $s = 0.37\pm 0.03$ for different detunings $\Delta$. First, we focus on the phase of the squeezing, i.e. the phase of the entangled two-photon wavefunction $\varphi_N(\tau) = \mathrm{Arg}\lbrace \phi_N(\tau) \rbrace$ averaged over $\Delta f$.
As in Fig.~\ref{fig_theta_plot}, we fit the averaged noise and extract the offset angle $\varphi$. Since the bandwidth $\Delta f$ contains the relevant frequencies, this method is equivalent to an integration over all frequencies and we introduce the frequency integrated entangled two-photon wavefunction $ 2 \pi \phi_N(\tau = 0) = \int_{-\infty}^{\infty} \phi_N(\omega) \mathrm{d}\omega$.
Figure~\ref{fig_phase_detuning} (a) shows the experimental values for  $\varphi_N(\tau=0)$ together with its theoretical prediction. 

In order to reconstruct the complex-valued function $\phi_N(\omega) = |\phi_N(\omega)| e^{i \varphi_N(\omega)}$, we access $\varphi_N(\omega)$ by fitting the phase in each frequency range individually. Figure.~\ref{fig_phase_detuning} (b) shows the phase of $\phi_N(\omega)$ as a function of $\omega$ together with its theoretical prediction for different laser-atom detunings $\Delta$. From the phase and the magnitude in frequency space, one can equivalently obtain the complex time-domain wavefunction  $\phi_N(\tau)$ by an inverse Fourier transform. Using this method, we obtain the experimental $\phi_N(\tau)$, and  Fig.~\ref{fig_phase_detuning} (c) shows three examples together with the corresponding theoretical prediction which does not contain any fit-parameter.

In conclusion, we observed the generation of squeezed light by sending weak coherent laser light through an ensemble of atoms weakly coupled to a nanofiber. The squeezing spectrum obtained via homodyne measurement gives direct access to the relative phase and magnitude of the two-photon wavefunction at the fiber output, allowing us to reconstruct the time-dependent two-photon wave function of the transmitted light. These measurements reveal the change in phase and magnitude of the entangled photons for different detunings and relate these to the squeezing spectrum. 

A recent theoretical proposal has suggested that correlation measurements can be used to reconstruct the scattering matrix of an arbitrary quantum scatterer \cite{ramosMultiphotonScatteringTomography2017}, and a first experimental step has been taken for a single quantum emitter \cite{jeannicExperimentalReconstructionFewphoton2020}. While we have focused on reconstructing the two-photon wavefunction, our measurement could also be extended to reconstruct the entire scattering matrix of an atomic ensemble.  Finally, while in this work we have focused on squeezing measurements, which quantify second-order amplitude correlations, studying higher-order correlations is also possible with our approach. For example, a non-zero third-order moment can unveil the expected non-Gaussian nature of the output photons and can also uncover the existence of three-body entanglement.

We are grateful to  A. S{\o}rensen, L. Orozco, P. Solano, J.-H. Mueller, B. Hacker and M. Kraft for stimulating discussions and helpful comments. We thank S. Rind for his support in building the experiment. 
We acknowledge financial support by the Alexander von Humboldt Foundation, the European Commission under the projects ErBeStA (No. 800942) and the ERC grant NanoQuaNt, and by the Austrian Science Fund (DK CoQuS Project No. W 1210-N16). M.S. acknowledges support by the European Commission (Marie Curie IF Grant No. 896957). S.M. and K.H. acknowledge funding from DFG through CRC 1227 DQ-mat, projects A05 and A06, and ``Nieders\"{a}chsisches Vorab'' through the ``Quantum-and Nano-Metrology (QUANOMET)''.

\bibliography{Squeezing}




\pagebreak
\widetext
\begin{center}
\vspace{1cm}
\textbf{\large Supplemental Material:  Unraveling two-photon entanglement via the squeezing spectrum of light traveling through nanofiber-coupled atoms  }
\end{center}
\setcounter{equation}{0}
\setcounter{figure}{0}
\setcounter{table}{0}

\makeatletter
\renewcommand{\theequation}{S\arabic{equation}}
\renewcommand{\thefigure}{S\arabic{figure}}
\renewcommand{\thesection}{S\arabic{section}}

\section{Theory calculation for the squeezing spectrum}

In this section we calculate the squeezing spectrum $S_{\theta}(\omega)$ in the low saturation limit~\footnote{The factor 8 is due to a different definition of $s$  compared to~\cite{mahmoodian_strongly_2018}.} ($s/8 \ll 1$). 
We start by calculating $\left\langle : \Delta\hat{X}_{\theta}(0) \Delta\hat{X}_{\theta}(\tau) :\right \rangle $
\begin{equation}
\left\langle : \Delta\hat{X}_{\theta}(0) \Delta\hat{X}_{\theta}(\tau) :\right \rangle = 
\frac{1}{4} e^{2i\theta} \left\lbrace  \left\langle \hat{a}(\tau)\hat{a}(0)\right\rangle  - \left\langle \hat{a}(\tau)\right\rangle \left\langle\hat{a}(0)\right\rangle    \right\rbrace + c.c. + 
\frac{1}{4} \left\lbrace \left\langle \hat{a}^{\dagger}(\tau)\hat{a}(0)\right\rangle  - \left\langle \hat{a}^{\dagger}(\tau)\right\rangle \left\langle\hat{a}(0)\right\rangle    \right\rbrace + c.c. .
\end{equation}
We assume a weak drive such that the dynamics of the system are captured by the one- and two-photon Fock states.
We calculate the expectation values for the output state $\ket{\mathrm{out}}$ which is a combination of the the one- and two-photon components, $\ket{\mathrm{out}}_1$ and  $\ket{\mathrm{out}}_2$, respectively. 
The respective definitions are given in~\cite{mahmoodian_strongly_2018} and from these we calculate to first order in $P_{\mathrm{in}}/P_{\mathrm{sat}}$
\begin{equation}
\bra{\mathrm{out}}\hat{a}(x)\hat{a}(0)\ket{\mathrm{out}} = e^{-|\alpha|^2/2} \bra{0}\hat{a}(x)\hat{a}(0)\ket{\mathrm{out}}_2 \simeq P_{\mathrm{in}} \psi_2(0,x).
\end{equation}
Here, we assumed a continuous coherent input state with power, in photons per second, $P_{\mathrm{in}} = \alpha^2 v_g/L$, where $\alpha$ is the coherent state amplitude and $L$ is a quantization length. We take the limit of $\alpha, L \rightarrow \infty$ such that the power of the coherent state approaches a finite constant. The two photon wave function $\psi_2(0,x)$ is defined as in Ref.~\cite{mahmoodian_strongly_2018}. To first order in $P_{\mathrm{in}}$ the term  $\left\langle \hat{a}^{\dagger}(\tau)\hat{a}(0)\right\rangle  - \left\langle \hat{a}^{\dagger}(\tau)\right\rangle \left\langle\hat{a}(0)\right\rangle $ vanishes and the term $\langle \hat{a}\rangle^2$ is given by $t^{2N}_{\Delta}P_{\mathrm{in}}$. This leads to the expression:
\begin{equation}
\begin{split}
\left\langle : \Delta\hat{X}_{\theta}(0) \Delta\hat{X}_{\theta}(x) :\right \rangle &= \frac{1}{4} P_{\mathrm{in}} e^{2 i \theta}  \left[\psi_2(0,x) - t_{\Delta}^{2N}\right] + c.c\\
&=  - \frac{\Gamma_{\mathrm{tot}}}{16 \beta} \frac{P_{\mathrm{in}}}{P_{\mathrm{sat}}} \operatorname{Re}{\left[ e^{2 i \theta}  \phi_N(x) \right]}
\end{split}
\end{equation}
where again we follow the definitions and notations of~\cite{mahmoodian_strongly_2018} for $\phi_N(x)$. Finally, for the squeezing spectrum, we obtain:
\begin{equation}\label{eq_full_S_omega}
{S}_{\theta}(\omega) = - \frac{\Gamma_{\mathrm{tot}}}{16 \beta}\frac{P_{\mathrm{in}}}{P_{\mathrm{sat}}}  \operatorname{Re}{\left[ e^{2 i \theta} \phi_N(\omega) \right]}.
\end{equation}
We note that for a resonant probe field $\phi_N(x)$ becomes entirely real.

The full expression of  $\phi_N(\omega)$ is given in \cite{mahmoodian_strongly_2018}. In the following we will focus on a resonant probe field  where in two limits simple analytic results can be obtained.
Furthermore, the two different shapes of the squeezing spectrum in those two limits can be directly seen in Fig.~\ref{fig_spec_asymptotic}.

\subsection{Small optical depth limit}
\begin{figure}[htbp]
\centering
\includegraphics[width=\textwidth]{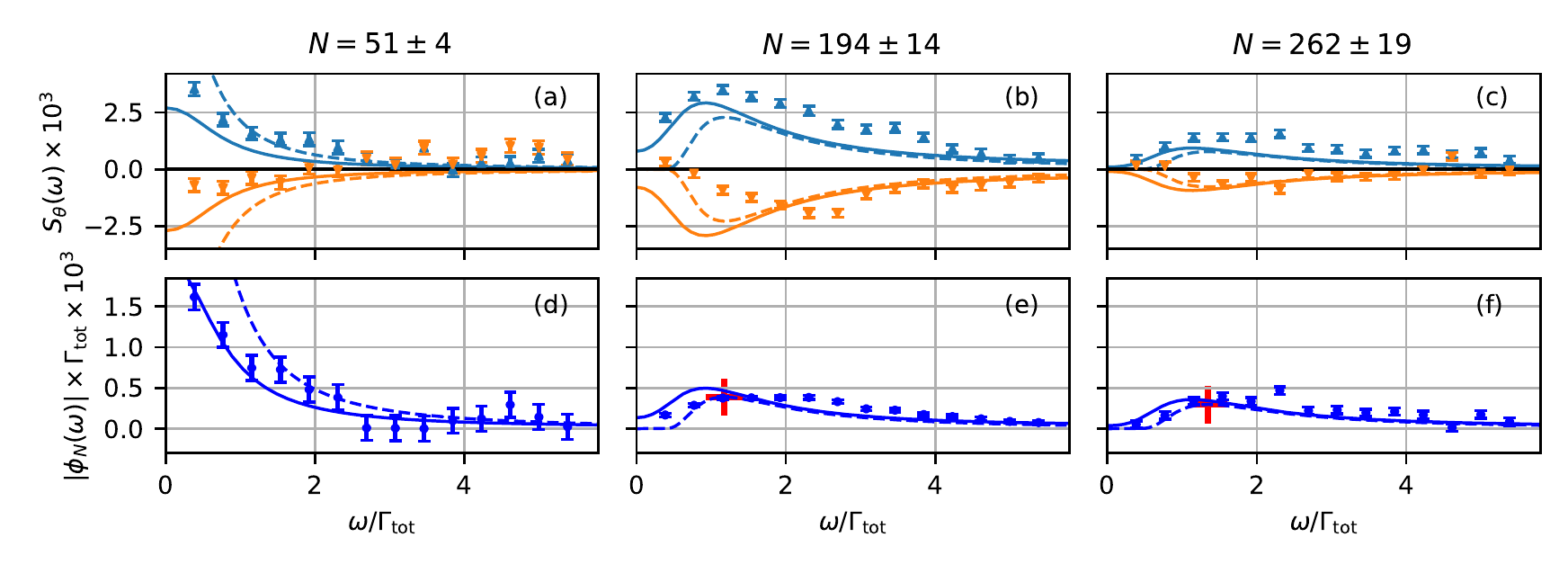}
\caption{Same as Fig.~\ref{fig_spectrum} together with the corresponding asymptotic expressions in dashed lines: (a) \& (d) with $N=51\pm4$ is close to the small optical depth limit in which the squeezing/two-photon spectrum is close to a Lorentzian. (b) \& (e),  (c) \& (f) with $N=194\pm14$ and $N=262\pm19$ respectively are close to the large optical depth limit where sidebands develop. The maximum squeezing as expected in the asymptotic limit and given by Eq.~\eqref{eq_maxima_sideband} is shown by the red cross.}
\label{fig_spec_asymptotic}
\end{figure}

In the low optical depth limit $\beta N \ll 1$ the squeezing is dominated by a coherent enhancement with the number of emitters. The entangled part of the two-photon wavefunction can be calculated  as:
\begin{equation}
\phi_N(x) = 4 N \beta^2 e^{-\frac{\Gamma_{\mathrm{tot}}}{2} |x|},
\end{equation}
which leads together with Eq.~\eqref{eq_full_S_omega} to:
\begin{equation}\label{eq_low_opt_depth}
{S}_{\theta}(\omega) = - N \beta \cos(2\theta) \frac{P_{\mathrm{in}}}{P_{\mathrm{sat}}} \frac{\Gamma_{\mathrm{tot}}^2}{\Gamma_{\mathrm{tot}}^2 + 4\omega^2}. 
\end{equation}
This is a Lorentzian and equivalent to the squeezing from $N$ single emitters. The squeezing peaks at $\omega = 0$, i.e. maximum squeezing is expected on resonance. The spectrum described by Eq.~\eqref{eq_low_opt_depth} is shown in Fig.~\ref{fig_spec_asymptotic} (a) and (d) as dashed lines.

\subsection{Large optical depth limit}
In the large optical depth limit $\xi_N^2 \equiv \beta N(1- \beta) \gg 1$ we have
\begin{equation}
\phi_N(\omega) =\frac{\beta \Gamma_{\mathrm{tot}}}{\omega^2} e^{-\frac{\xi_N^2 \Gamma_{\mathrm{tot}}}{\omega^2}}.
\end{equation}
This leads together with Eq.~\eqref{eq_full_S_omega} to:
\begin{equation}
{S}_{\theta}(\omega) = - \frac{1}{16} \cos(2\theta)\frac{P_{\mathrm{in}}}{P_{\mathrm{sat}}} \frac{\Gamma_{\mathrm{tot}}^2}{\omega^2}e^{-\frac{\xi_N^2 \Gamma_{\mathrm{tot}}}{\omega^2}}.
\end{equation}
This yields maximum squeezing at a frequency $\omega_\mathrm{max} = \pm \xi_N \Gamma_{\mathrm{tot}}$ which is shown in Fig.~\ref{fig_spec_asymptotic} by the red cross.  The peak amplitude is given by
\begin{equation}
S_\theta ( \pm \xi_N \Gamma_{\mathrm{tot}} ) = - \frac{1}{16} \cos (2\theta) \frac{P_{\mathrm{in}}}{P_{\mathrm{sat}}} \frac{1}{e \xi_N^2},\label{eq_maxima_sideband}
\end{equation}
i.e. the amplitude is proportional to $1/N$. This asymptotic  prediction is shown in Fig.~\ref{fig_spec_asymptotic} (b) \& (e) and (c) \& (f) where we have $\xi_N^2= 1.4 $ and $\xi_N^2= 1.8 $.

\subsection{Evolution of the entangled photon pairs along the fiber}
\begin{figure}[h!]
    \centering
    \includegraphics[width=0.5\linewidth]{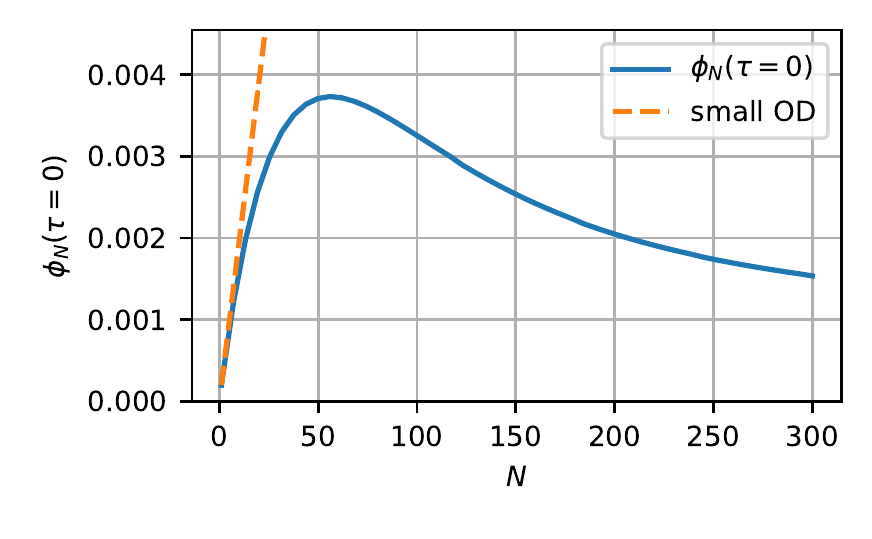}
    \caption{The evolution of $\phi_N(\tau=0)$ in blue as a function of $N$ with $\beta =0.007$ as in the experiment. The dashed line corresponds to the small optical depths approximation $\phi_N(\tau=0)= 4\beta^2 N $ which neglects absorption and reflects the coherent build up of the squeezed photons.}
    \label{fig_phi_N_0}
\end{figure}
The squeezing -- or the entangled photon pairs -- can be characterized by $\phi_N(\tau=0) $ which is the entangled wavefunction integrated over all frequencies    $\phi_N(\tau=0)  = 1/(2\pi) \int_{-\infty}^{\infty}\phi_N(\omega) \mathrm{d}\omega$ as shown in Fig.~\ref{fig_phi_N_0}. For small optical depths absorption of the entangled pairs can be neglected. The photon pairs at each atom are scattered with the same phase which leads to a $N$-times enhanced creation of entangled photon pairs in the wavefunction due to a coherent build up in forward scattering. For larger ensembles, the absorption of the photon pairs along the chain cannot be neglected anymore which manifests itself in a reduced amplitude of $\phi_N(\tau=0)$. For large OD, $\phi_N(\tau=0)$ decreases. 

\section{The setup}

\subsection{Measurement of $\beta$}
\begin{figure}[htb]
\centering
\includegraphics[width=0.5\textwidth]{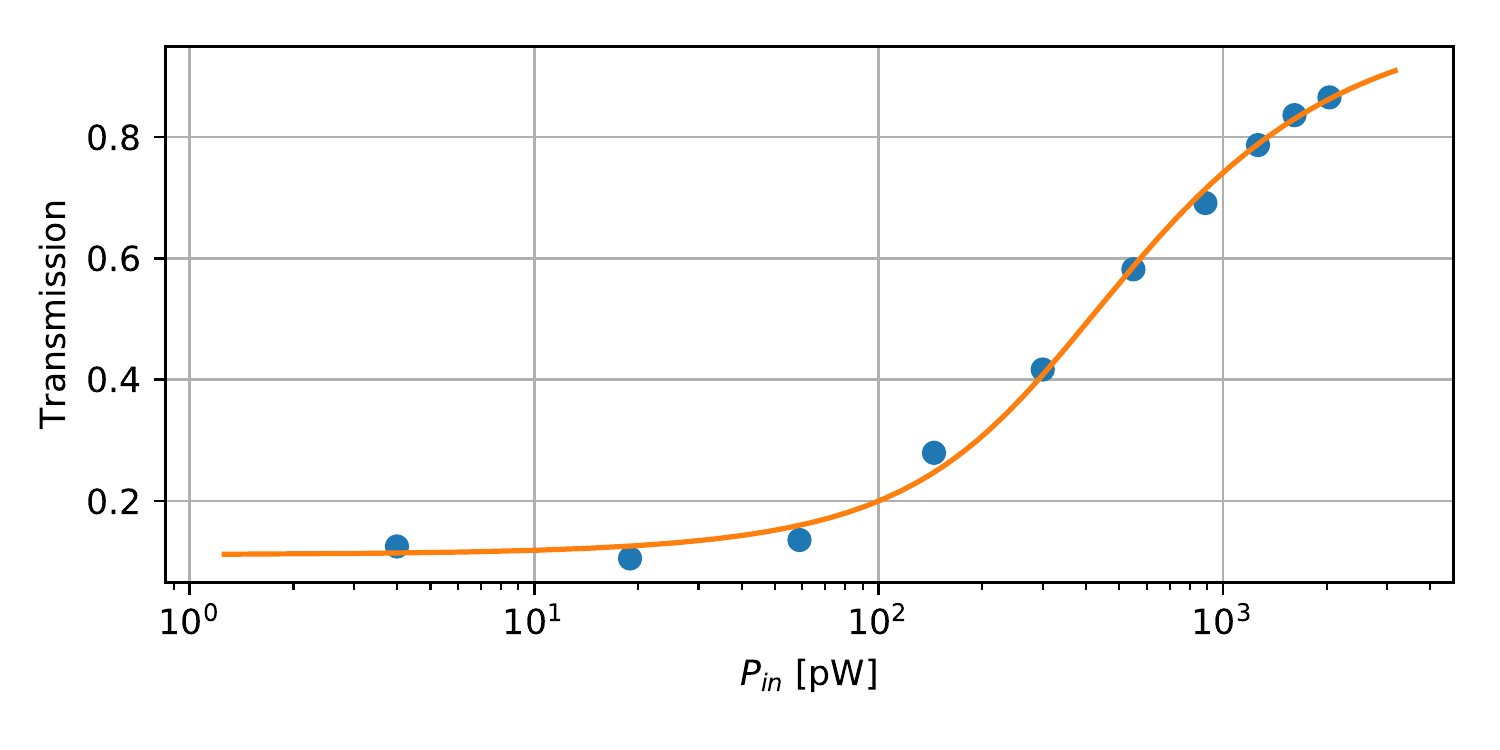}
\caption{Ensemble transmission as function of input power $P_{\mathrm{in}}$. The orange line is a fit of the data (blue dots). The free fitting parameters are $N$ and $\beta$.}
\label{fig_beta_measurement}
\end{figure}

In order to measure the coupling strength $\beta$ in our experiment we perform a transmission measurement for varying input power from $s=0.03$ to $s=15$ and note the corresponding transmission. The transmission $T$ is given by an extended Lambert-Beer law~\cite{stenholmFoundationsLaserSpectroscopy2012} and one can show that it follows
\begin{equation}
T = \frac{\mathfrak{W}\left( e^{-4 \beta N + \frac{P_{\mathrm{in}}}{P_{\mathrm{sat}}}} \frac{ P_{\mathrm{in}} }{ P_{\mathrm{sat}} }  \right) }{\frac{P_{\mathrm{in}}}{P_{\mathrm{sat}}}}
\end{equation}
where $\mathfrak{W}$ denotes the Lambert W function. The fitted experimental data is shown in Fig.~\ref{fig_beta_measurement}  from which we obtain $\beta = 0.0070(5)$.

\subsection{Measurement of the efficiency of the detection setup.}
In order to measure the efficiency of our homodyne setup, we characterize a well-defined coherent state without atoms.  We measure the power of the coherent state with a single photon detector and compare it to the power obtained from the homodyne setup. This leads to an overall photon detection efficiency of $\eta = 0.22$. This value is in agreement with the efficiency obtained from the optical losses, mode matching and electronic noise~\cite{lvovsky_squeezed_2014,appel_electronic_2007,bachor_hansa_quantum_2019}.

\subsection{Measurement of the squeezing spectrum}
In order to measure the normally ordered squeezing spectrum $S(\omega)$ -- the Fourier transform of the auto-correlation function of $\Delta \hat X(t)$ -- we first compute the Fourier spectrum of the current $I(t)$ via a Fast-Fourier transform. By using the Wiener-Khinchin theorem this quantity, normalized to the vacuum reference, is the non-normally order squeezing spectrum 
\begin{equation}
\tilde{S}_{\theta}(\omega) = 4 \int_{-\infty}^{\infty} \mathrm{d}\tau \left\langle  \Delta\hat{X}_{\theta}(0) \Delta\hat{X}_{\theta}(\tau) \right \rangle e^{i \omega \tau}  = \frac{|\int_{-\infty}^\infty I_{\theta}(t) e^{i\omega t}  \mathrm{d}t  |^2 }{|\int_{-\infty}^\infty  I_\mathrm{vac}(t) e^{i\omega t} \mathrm{d}t |^2 }\label{eq_spectrum_non_normal}.
\end{equation}
The differential detector current measured for a given $\theta$ is denoted as $I_{\theta}$ and $I_\mathrm{vac}(t)$ is recorded from the vacuum measurement, i.e. the case where the transmitted light from the nanofiber is blocked.
 The normalization to the vacuum reference eliminates the normalization ambiguities of the Fourier Transform. The prefactor $4$ in Eq.~\eqref{eq_spectrum_non_normal} is due to the factor $\frac{1}{2}$ in our definition of $\hat X_{\theta}(t)$.  We finally obtain the normally ordered squeezing spectrum via $ {S}_{\theta}(\omega) = \frac{1}{4}\left[\tilde{S}_{\theta}(\omega) - {1}\right] $.  
 A state of no noise is given by $\tilde{S}_{\theta}(\omega) = 0$ from which follows that the normally ordered quantity for no noise is ${S}_{\theta}(\omega) = -1/4$. For a coherent state we have $S_{\theta}(\omega) = 0$.

A state of no noise is given by $\tilde{S}_{\theta}(\omega) = 0$ from which follows that the normally ordered quantity for no noise is ${S}_{\theta}(\omega) = -1/4$.  The integrated normally ordered squeezing of no noise is  $\langle \normalorder{\hat X^2_\theta (\tau = 0)}\rangle/\Delta f = -1/2$ where $\Delta f = f_\mathrm{max} - f_\mathrm{min}$.
 
\subsection{The cycling transition}
The measurements were all carried out on the $6S_{1/2}, F=4 \rightarrow 6P_{3/2}, F'=5$ transition (D2-line). Initially, the atoms are in a combination of all magnetic quantum states $m_F$ states. Before the homodyne measurement, the atoms are optically pumped for a few microseconds by resonant light in zero magnetic field~\cite{mitschExploitingLocalPolarization2014}. This transfers the atoms to the cycling transition $F=4,m_F=4 \rightarrow F=5,m_F = 5$ of the Cesium D2-line. Then, the system behaves as an effective two-level system.

\section{Phase and amplitude squeezing}
\begin{figure}[h!]
    \centering
    \includegraphics[width=0.3\linewidth]{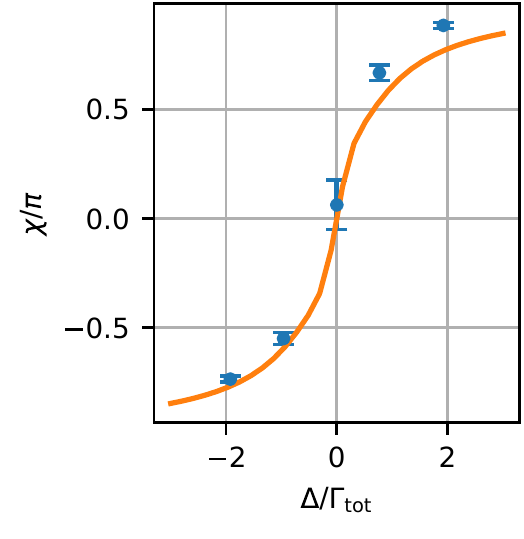}
    \caption{The measured squeezing angle $\chi$ as a function of input detuning $\Delta$ (blue data points) together with the corresponding theoretical prediction (solid line) for measured $N = 140\pm 10$ atoms.}
    \label{fig_squeezing_angle}
\end{figure}

In the main text, we investigate the relative phase of the entangled two-photon wavefunction $\varphi_N(\tau)  = \mathrm{Arg}\lbrace \phi_N(\tau) \rbrace$ with respect to the input light. For off-resonant light, the atoms imprint a phase-shift on the light as given by Eq.~\eqref{eq_phase_shift_atoms}. Instead of investigating $\varphi_N(\tau)$, one can equivalently study the squeezing angle with respect to the output light field $\langle X_\theta(t)\rangle$. For this, we introduce the squeezing angle $\chi = \varphi_N(\tau=0) - \mathrm{Arg}\lbrace t_{\Delta}^N\rbrace$. For $\chi=0$, the quadrature fluctuations $\langle \normalorder{\Delta X_\theta^2(0)}\rangle$ and the output light field $\langle X_\theta(t)\rangle$ are in phase and the light is amplitude squeezed. For $\chi = \pm \pi$, the light is phase squeezed. Figure~\ref{fig_squeezing_angle} shows the measured angle $\chi$ as function of detuning $\Delta$ together with its theoretical prediction. 

We would like to point out, that in contrast to the amplitude of the entangled part of the two-photon wavefunction, the measurements of the angles $\chi$, as well as $\varphi_N$ do not depend on the detection efficiency $\eta$ in our setup.

\section{Squeezing from untrapped atoms}
\begin{figure}[htbp]
\centering
\includegraphics[width=0.5\linewidth]{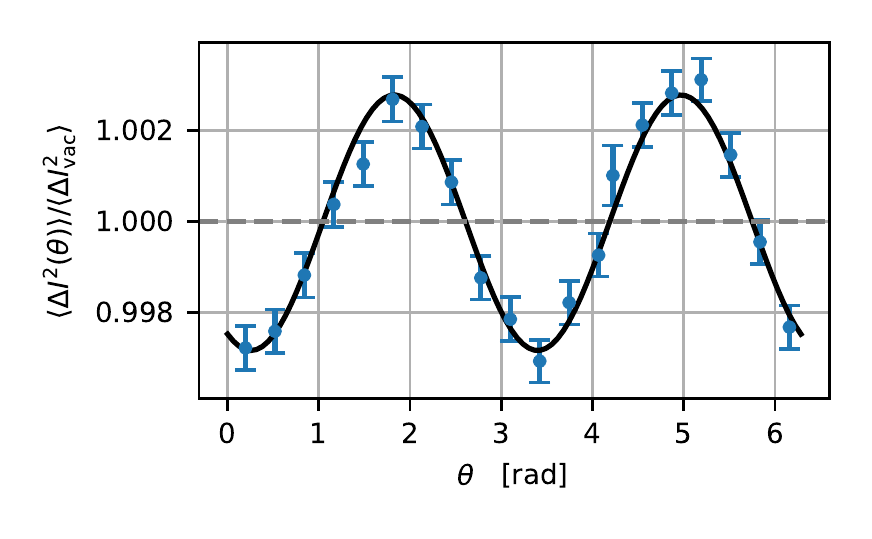}
\caption{The squeezing from untrapped atoms in the bandwidth from $f_{\mathrm{min}} = \unit[1.5]{MHz}$ to $f_{\mathrm{max}} = \unit[23]{MHz}$ with an input power of $P_{\mathrm{in}} = \unit[20]{pW}$.}
\label{fig_squeezign_MOT}
\end{figure}

It is interesting to  note that the underlying physics does not depend on the periodicity of the lattice potential that is introduced by the standing wave of the red detuned trap fields. As confirmation, we performed a squeezing measurement without the optical lattice which could potentially introduce a periodicity. Similar to the case where atoms are trapped we observe a squeezing of $0.31\% \pm 0.05 $ within the same bandwidth from $f_{\mathrm{min}} = \unit[1.5]{MHz}$ to $f_{\mathrm{max}} = \unit[23]{MHz}$. This agrees with our expectations as the lattice does not significantly change the average $\beta$-factor of the atoms. Figure~\ref{fig_squeezign_MOT} shows the normalized homodyne detector noise $\langle \Delta I^2(\theta)\rangle/ \langle \Delta I^2_{\mathrm{vac}}\rangle$ as a function of $\theta$ which shows that this configuration also exhibits squeezing.

\end{document}